\documentclass[
aps, pre,
twocolumn, 
showpacs, 
nopreprintnumbers, 
amsmath,amssymb,
letterpaper
]{revtex4}

\usepackage{graphicx}
\usepackage{dcolumn}
\usepackage{bm}

\begin{document}


\title{ 
Approximating the ground state of fermion system by multiple determinant states:
matching pursuit approach
}

\author{Quanlin \surname{Jie}}
\email[E-mail: ]{qljie@whu.edu.cn}
\affiliation{%
Department of Physics, Wuhan University,
Wuhan 430072, P. R. China
}%

\date{\today}

\begin{abstract} 
We present a simple and stable numerical method to approximate the ground
state of a quantum many-body system by multiple determinant states. This
method searches these determinant states one by one according to the matching
pursuit algorithm. The first determinant state is identical to that of the
Hartree-Fock theory. Calculations for two-dimensional Hubbard model serve as a
demonstration.
\end{abstract}

\maketitle

Searching a single determinant state to approximate a quantum ground state,
namely, the Hartree-Fock (HF) algorithm,  plays important role in the
understanding of nuclear, atomic, and molecular structures. It is a long
standing effort to extend the HF theory into a truly first principle method by
searching multiple determinant states to span a quantum state, for recent
examples, see~\cite{1,2,10a,18,7} and references therein. The attracting
feature is that this approach is very stable and free from the sign problem.
It in principle can apply to a wide variety of systems.  However, first
principle calculation in terms of multiple determinant states is still a
challenge. In fact, including multiple determinant states in the variational
treatment, which is the common approach, often results in very complicated
formulations.  The computation cost is usually impractically demanding.  Some
realistic implementations impose restrictions on the determinant states.  For
example, the Multi-configuration Hartree-Fock theory~\cite{1}, a time
dependent extension to the HF theory, requires single particle states to be
orthogonal with each others. Here, we use a new approach to approximate the
quantum ground state via multiple determinant states.

Here, based on the matching pursuit (MP) algorithm~\cite{3,4,5,6}, we show a
numerical method to search determinant states to span the ground state of a
fermion system. The determinant states are found one by one from all possible
determinant states.  Searching the first determinant state is identical to the
Hartree-Fock theory. A significant feature of the current method is that
several tens of basis determinant states are enough for reasonable result, and
one can reach high accuracy by searching one or two thousands basis states.
These numbers of basis states are several orders smaller than that of the
Stochastic diagonalization algorithm~\cite{7}, which searches orthogonal
determinant basis states stochastically to span a quantum wave function.  In
comparison with other algorithms of search determinant states to span a
fermion ground state, such as the Path-Integral Renormalization Group (PIRG)
algorithm~\cite{10a,18}, the current MP based method is quite simple and
efficient.

The MP algorithm is originally designed for signal processing~\cite{3}. It is
now popular on the engineering community for coding, analysis, and compression
of video and audio data~\cite{4,5,6}. This algorithm searches some basis
states from an over-complete basis set to represent a sequence of data.  The
basis states are found one by one. The convergence of the MP algorithm is
proved mathematically. For sufficient redundancy of the over-complete basis
set, the convergence can be exponential~\cite{6}.  The MP algorithm is
insensible to the dimension of the data, and thus promises applications in
quantum many-body systems. In Ref.~\cite{8}, the authors employ this algorithm
to propagate quantum wave functions via split operator method in the Gaussian
wave packet basis. A encouraging result is that several tens of Gaussian wave
packets are able to accurately represent quantum wave function of a
20-dimensional model.

The goal of MP algorithm is to obtain a sparse representation of a signal. To
represent a quantum many-body wave function, $\psi$, the MP algorithm searches
an over-complete basis set and finds some basis states, $\phi_1$, $\phi_2$,
$\cdots$, $\phi_n$, such that the combination of the basis states,
$\psi_n=\alpha_1\phi_1+\cdots+\alpha_n\phi_n$ can best approach the state
$\psi$.  Mathematically, that is to require $\psi_n$ has minimum distance with
$\psi$, i.e. $|\psi-\psi_n|$ reaches minimum. The basis states are found one
by one.  At $k$-th step, the basis state $\phi_k$ is obtained such that the
combination of the basis states $\psi_k=\alpha_1\phi_1+\cdots+\alpha_k\phi_k$
has minimum distance with the state $\psi$, i.e. $|\psi-\psi_k|$ has minimum
for all possible choice of $\phi_k$. Each more step brings the $\psi_k$ closer
to the target state $\psi$, i.e., the distance $|\psi-\psi_k|$ decreases with
$k$.

The eigenvalue problem is equivalent to find minimum values of the Rayleigh
quotient 
\begin{equation}\label{eq1}
E={\langle \psi|H|\psi\rangle}/{\langle \psi|\psi\rangle},
\end{equation} 
where $H$ is the Hamiltonian and $\psi$ is the trial wave function.
Calculation of the ground state by the MP algorithm is to search some basis
states to span the ground state. The basis states are found one by one from an
over-complete basis set. Each searching process obtains one basis state such
that the combination of this basis state and those already found ones
minimizes the Rayleigh quotient for all possible choice of the current basis
state. This process of finding a new basis state continues until convergence
of $E$.  Without loss of generality, in the following discussions, we focus on
fermion systems, and use all possible Slater determinant states as
over-complete basis set.

Note that, for fermion systems, the first step is to find a Slater determinant
state that minimizes the Rayleigh quotient. This is just the well known
Hartree-Fock approximation.  We employ an iterative method to search a
new determinant state, including the first one. We denote the single particle
basis states as $|i\rangle$, ($i=1,\cdots,n$), and $a^+_i$ ($a_i$) the
operator for creation (annihilation) of the state $|i\rangle$, i.e.,
$|i\rangle=a^+_i|0\rangle$ with $|0\rangle$ the vacuum state. A determinant
state can be expressed as
\begin{equation}\label{eq2} 
|\phi\rangle=\prod_{j=1}^m F^+_{j}|0\rangle, 
\end{equation}
where $m$ is particle number and $F^+_{j}$ ($F_{j}$) is creation
(annihilation) operator for single particle state,
$F^+_{j}=c_{1j}a^+_1+\cdots+c_{nj}a^+_n$. Searching for the determinant state
$|\phi\rangle$ is equivalent to find the coefficients $\{c_{ij}\}$ (or the
operators $\{F^+_j\}$).

We use an iterative relaxation procedure to search the operators $\{F^+_j\}$.
From an initial trial state in the form of (\ref{eq2}) which can be chosen
randomly, we optimize $F^+_1$, $F^+_2$, $\cdots$, $F^+_m$ consecutively. Each
step of the optimization lowers the Rayleigh quotient. This iteration
continues until the convergence of the Rayleigh quotient. Note that the
determinant state (\ref{eq2}) is a multi-linear function of the coefficients
$\{c_{ij}\}$. For a fixed $j$, $|\phi\rangle$ is just a linear function of
$c_{1j}$, $\cdots$, $c_{nj}$:
\begin{equation}\label{eq3}
|\phi\rangle = \sum_i c_{ij} |\phi_{ij}\rangle,
\end{equation}
where $|\phi_{ij}\rangle = {\partial |\phi\rangle}/{\partial c_{ij}}$.  Thus
an approximate ground state $\Psi_k=\sum_i \alpha_i\phi^{(i)} + \alpha\phi$
can be written as 
\begin{equation}
\Psi_k=\sum_i \alpha_i\phi^{(i)} + \sum_i \alpha
c_{ij}\phi_{ij}.
\end{equation}  
This means that we can improve $\Psi_k$ and hence update
the operator $F^+_j$ by finding the lowest eigenstate of the Hamiltonian in
the subspace spanned by $\{|\phi_{ij}\rangle$, $i=1,\cdots,n\}$ and those
previously found determinant states $|\phi^{(i)}\rangle$.

Such relaxation procedure to update the operators $F^+_j$ is the key
ingredient of this contribution. Suppose we have already obtained $k-1$
determinant states $|\phi^{(i)}\rangle$, the searching process for $k$-th
determinant state $|\phi^{(k)}\rangle=|\phi\rangle$ in the form (\ref{eq2})
involves the following iteration:

(1) randomly generate a determinant state $|\phi\rangle$.

(2) For $j=1,2,\cdots,m$, do the following iteration loop to update
$|\phi\rangle$:

(2a) Calculate the matrix elements of the Hamiltonian in the subspace
$\Xi^{(k)}_j$ spanned by $\{|\phi^{(1)}\rangle, \cdots, |\phi^{(k-1)}\rangle$,
$|\phi_{1j}\rangle, \cdots, |\phi_{nj}\rangle\}$;

(2b) Find the ground state $\Psi_k^{j}$ of the Hamiltonian in the above
subspace $\Xi^{(k)}_j$, $\Psi_k^{j}=\sum_i \alpha_i\phi^{(i)} + \sum_i
\beta_{ij}^k\phi_{ij}$;

(2c) Update $F^+_j$ by  setting $c_{ij}=\beta^{k}_{ij}$ ($i=1,\cdots,n$); Then
make $F^+_j|0\rangle$ orthogonal to other single particle states
$\{F^+_l|0\rangle,\ l\neq j\}$, and restore $F^+_j|0\rangle$ to unit length
by a normalization procedure.

(3) Check the convergence of the Rayleigh quotient. Repeat the step (2) until
reaching convergence.

In case of $k=1$, the above searching process of finding the first determinant
state is the same as the Hartree-Fock algorithm. A randomly generated initial
trial state needs several tens of iteration rounds to converge. Each of
subsequent determinant states needs about similar rounds of iteration.  Here,
the main numeric cost is the step (2a) for calculation of the matrix elements
of the Hamiltonian between basis states. The step (2b) of finding lowest
eigenstate in the subspace can be implemented efficiently via iteration
algorithm~\cite{9,10} that needs only small portion of the computation cost.

Starting from $k=2$, the number of iteration to obtain $\phi_k$ depends on the
initial choice. If a trial state has large overlap with the state
$(H-E_{k-1})|\Psi_{k-1}\rangle$, one may reach convergence by just a few
rounds of iteration. Here, $E_{k-1}$ and $\Psi_{k-1} = \sum_{i=1}^{k-1}
\alpha_i\phi^{(i)}$ are the approximate ground state energy and wave function
obtained in the previous step.  One can understand this property by
considering minimization of the Rayleigh quotient in the two dimensional
subspace spanned by $\Psi_{k-1}$ and the trial state $\phi$~\cite{7}. From
this observation, we perform a preparing treatment of the trial state before
step (2) of the above iteration procedure.

The preparing treatment of the initial trial state $\phi$ is to modify the
state $\phi$ so that it has maximum overlap with the state
$(H-E_{k-1})|\Psi_{k-1}\rangle$.  This procedure is easy to carry out by
exploiting the fact that state $\phi$, or the overlap $\langle \phi
|H-E_{k-1}| \Psi_{k-1}\rangle$, is a multi-linear function of the coefficients
$c_{ij}$ (or the operators $F^+_j$). We maximize the overlap iteratively by
updating the operators $F^+_j$, $(j=1,\cdots,m)$, consecutively. Usually, 3 to
5 rounds of the iteration are enough. After such preparing treatment, one
usually needs about 2 to 3 iteration rounds of the searching process to
minimize the Rayleigh quotient. Thus, such preparing treatment makes the
overall procedure about 5 to 10 times faster. 

As an optional choice to achieve high accuracy, one can perform backward
optimization after reaching convergence in the above procedure. This procedure
updates the already found basis states one by one (One can also choose to
update some selected basis states~\cite{4}). The operation to update a basis
state is the same as searching a new basis state. It is numerically expansive
to perform the backward optimization. In fact, searching the basis states one
by one is a kind of restriction on the determinant states, and the backward
optimization means removing such constraint.

At first sight, the current method shares some features with the Path-Integral
Renormalization Group (PIRG) algorithm~\cite{10a,18}. However, based on
different strategies, the PIRG and the current method are two different
methods of searching basis determinant states.  The PIRG filters out the
ground state by repeatedly expanding $e^{-\tau H}|\psi\rangle$ into summation
of determinant states and keeping some of the determinant states as new basis
states to update the trial ground state $|\psi\rangle$. At each step, the PIRG
must update whole basis states, while the current method only adds (or
updates) one basis state via relaxation method and exploiting the
multi-linearity of the determinant states.  Updating basis states in PIRG,
i.e., choosing some determinant states from those ones that span the state
$e^{-\tau H}|\psi\rangle$, involves diagonalization of many sizable matrices.
The diagonalization in subspace is a major numeric cost of PIRG, while it
takes only a small portion of numeric operations in current method. The
current method only calculates matrix elements of the Hamiltonian
between determinant states, this is much easier and more efficient than
expanding $e^{-\tau H}|\psi\rangle$ (or $H|\psi\rangle$) into summation of
determinant states.

We test the above method via the two dimensional fermionic Hubbard mode on a
$N=L\times L$ square lattice with periodic condition. The Hamiltonian reads
\begin{equation}\label{eq4} 
H=-t\sum_{\langle
ij\rangle}(c^+_{i\sigma}c_{j\sigma}+c^+_{j\sigma}c_{i\sigma}) +U\sum_i
n_{i\uparrow}n_{i\downarrow}.  
\end{equation} 
Here $c^+_{j\sigma}$ ($c_{j\sigma}$) is the creation (annihilation) operator
of an electron with spin $\sigma$ at $j$-th site and $n_{j\sigma} =
c^+_{j\sigma} c_{j\sigma}$.  $U$ is the on-site Coulomb energy. The summation
$\langle ij\rangle$ runs over nearest-neighbor sites.

\begin{table}[h]
\caption{ Ground state energies of some the $4\times 4$ systems.
}\label{tab1}
\begin{tabular}[t]{lllllll}
\hline\hline 
system    & $U/t$ &  CPMC     & SD     & MP        & N   & Exact         \\
\hline 
$10/16$   & 4     &  -19.5808 &-19.58  & -19.5775  &1874 & -19.5808      \\
$14/16$   & 4     &  -15.7296 &-15.49  & -15.7107  &1865 & -15.7446      \\
$16/16$   & 4     &           &-13.59  & -13.5963  &1622 & -13.6219      \\
$10/16$   & 8     &  -17.4800 &-17.40  & -17.4625  &2000 & -17.5104      \\
$14/16$   & 8     &  -11.648  &        & -11.6763  &2000 & -11.8688      \\
$16/16$   & 8     &           &        & -8.41565  &1850 & -8.46889      \\
$14/16$   & 12    &  -9.696   &        & -9.79500  &2000 & -10.0515      \\
$16/16$   & 12    &           &        & -5.95238  &1950 & -5.99222      \\
\hline 
\hline 
\end{tabular}
\end{table}
Table \ref{tab1} shows ground state energies (in the unit of $t$) of some
$4\times4$ systems.  The column ``system'' indicates the number of electrons
versus the lattice number. $N$ is number of basis determinant states to span
the ground state of the current method (MP).  We list the results of the
Constrained Path Quantum Monte Carlo (CPMC)~\cite{11}, Stochastic
Diagonalization (SD)~\cite{7}, and exact diagonalization~\cite{12,13,14,15}
for comparison.  The accuracy of our method is almost unchanged for various
interaction strength $U$ and filling number. This demonstrates the stability
of the method. All initial trial states for searching basis determinant states
are randomly generated without any symmetry consideration.  We use convergence
rate $\epsilon=2|E_n-E_{n-1}|/|E_n+E_{n-1}|$ to determine the number of basis
states, where $E_n$ and $E_{n-1}$ are ground state energies obtained with $n$
and $n-1$ basis states, respectively. The searching for basis states stops if
$\epsilon$ is smaller than a criteria $\epsilon_0$, or maximum acceptable
number of basis states is reached. Usually, $\epsilon_0=10^{-5}$ is enough to
obtain quite reasonable result. At this setting, one usually needs several
hundreds basis states which increases slowly with $U$. If
$\epsilon_0=10^{-6}$, one needs several thousands of basis states for
convergence. Roughly speaking, result from about 100 basis states is quite
well. The interest point is that the beginning several tens of determinant
states, usually less than 60 basis states, make dominant contribution. And the
first one, i.e., the HF approximation, contributes most.  The number of
dominant basis states increases slowly with the interaction strength $U$.
After the dominant basis states, the contribution from each of following ones
drops rapidly.

\begin{table}[h]
\caption{Correlation functions of some the $4\times 4$ systems. 
}\label{tab2}
\begin{tabular}[t]{llllll}
\hline\hline 
Method  & system  & $U/t$ & $\rho(2,1)$ & $S_m(\pi,\pi)$ & $S_d(\pi,\pi)$ \\
\hline 
CPMC    & $10/16$ & 4     &  -0.0556    & 0.731           & 0.504          \\
MP(800) & $10/16$ & 4     &  -0.0559    & 0.7315          & 0.5089         \\
Exact   & $10/16$ & 4     &  -0.0556    & 0.73            & 0.506           \\
\hline 
QL      & $16/16$ & 4     &  -0.0363    & 3.42            & 0.3946          \\
MP(437) & $16/16$ & 4     &  -0.0478    & 3.490           & 0.3888          \\
Exact   & $16/16$ & 4     &  -0.0475    & 3.64            & 0.385           \\
\hline 
CPMC    & $10/16$ & 8     &  -0.0462    & 0.761           & 0.4403          \\
MP(800) & $10/16$ & 8     &  -0.0493    & 0.7645          & 0.4412          \\
Exact   & $10/16$ & 8     &  -0.0485    & 0.75            & 0.443           \\
\hline 
\hline 
\end{tabular}
\end{table}
Table \ref{tab2} shows ground state's correlation functions of some $4\times
4$ systems. The comparing results of CPMC, Quantum Langevin (QL), and exact
diagonalization are from~\cite{11}, \cite{15}, and \cite{12,15}, respectively.
Here $S_m$ and $S_d$ are magnetic and density structure factors~\cite{11},
respectively; and $\rho(\mathbf{r})$ is the one body density matrix. The
number in the column ``method'' is the number of basis states of our method
(MP). The current method obtains the ground state energy and wave function at
the same time. Then calculation of the correlation functions and other related
quantities is a trivial task that simply reads the wave function. In
comparison with the exact result, we see that the wave functions are almost in
the same accuracy as the correspondent energies. Again, the beginning several
tens of basis states make major contribution. To demonstrate this property, as
indicated in the parentheses, we use only several hundreds of basis states to
calculate the correlation functions in table \ref{tab2}. Since our program
does not perform any symmetry treatment, we can only compare non-degenerated
ground states with the exact result. In fact, the present method can take into
account of symmetries. After the searching process of finding  basis
determinant states, the resultant wave function is usually a combination of
ground states with different symmetries.  One may employ, e.g., the project
technique~\cite{16} to filter out the target symmetry. This may further
improve the accuracy.

\begin{table}[h] \caption{Ground state energies (in the unit of $t$) of some
large systems with $U=4t$.
}\label{tab3} \begin{tabular}[t]{llllll} \hline\hline system          & QMC &
VMC     & SD     & MP             & $N$      \\ \hline $26/6\times 6$  &
-42.32   &         &-40.77  & -41.0757        & 400      \\ $34/6\times 6$  &
-33.30   & -32.76  &        & -32.7323        & 940      \\ $36/6\times 6$  &
-30.96   & -30.384 &        & -30.5166        & 905      \\ $50/8\times 8$  &
-72.80   &         &-67.00  & -68.5029        & 630      \\ $54/8\times 8$  &
-67.55   &         &        & -63.8981        & 560      \\ $62/8\times 8$  &
-57.70   &         &        & -55.5255        & 490      \\ $64/8\times 8$  &
-55.23   &         &        & -53.583         & 472      \\
$100/10\times10$& -86.70   &         &        & -82.9549        & 152      \\
\hline \hline \end{tabular} \end{table}
Table \ref{tab3} shows ground state energies for large system size that exact
diagonalization is impossible.  Here $N$ is the number of basis states of the
current method (MP).  Our result is quite close to that of the SD~\cite{7},
and the variational quantum Monte-Carlo (VMC)~\cite{18}.  It is worth to note
that the number of basis states of our method is several orders smaller than
that of  the SD algorithm. As a consequence, our method needs much less
memory, and there is no need for external storage. There are several percent
of discrepancy with the Quantum Monte Carlo (QMC) result~\cite{17}, this
disagreement increases with the system size. This needs further
investigations. The discrepancy between QMC result and the strict variational
result is also found by other authors, see, e.g.,~\cite{10a,16,18}.  For
practical applications, the extrapolation method introduced in PIRG's
implementation is an useful tool to handle the discrepancy with QMC
results~\cite{10a,16}.  Similar to the $4\times4$ cases, the beginning several
tens of basis states make dominant contribution. For a fixed interaction
strength $U$, the contribution from the dominant basis states increases with
the system size.  This means that one needs less basis states for larger
system size. On the other hand, the computation cost to search a basis state
scales about quadratically with the system size. The role of preparing step
for searching a basis state is more significant for larger system size. With
out this preparing step, after several tens of dominant basis states, the
overlap between a randomly generated trial state $\phi$ and the state
$(H-E_{k-1})|\Psi_{k-1}\rangle$ almost vanishes. One must substantially
increase accuracy requirement for following steps, which in turn increases
numeric cost. The computation cost scales about quadratically with the number
of basis states.

\begin{figure}[h]
\includegraphics[angle=-90,width=\columnwidth,clip]{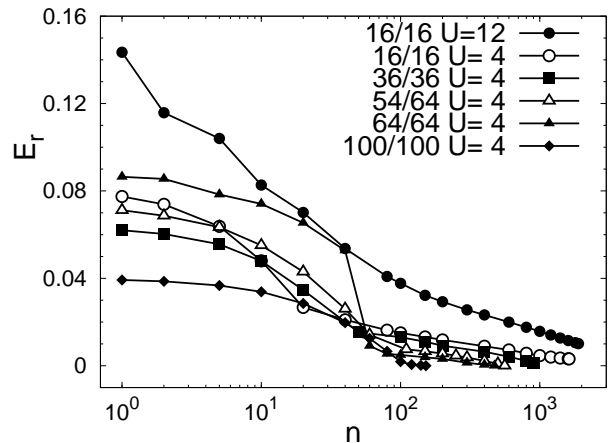}
\caption{\small
Relative error $E_r$ versus number of basis states $n$ for various filling
numbers and system sizes.
}\label{fig1} \end{figure}
Fig. \ref{fig1} shows the relative error $E_r=|(E_n-E_0)/E_0|$ versus the
number of basis states $n$, where $E_n$ is the ground state energy obtained
with $n$ basis states, and $E_0$ is the converged ground state energy (or
exact ground state energy if available).  This illustrates the overall
properties of the method. The first basis state makes most important
contribution. It accounts for about $80\%$ to $95\%$ of the ground state
energy depending on system size and correlation strength.  Roughly speaking,
the mean field effect increases with the system size, and decreases with the
correlation between electrons of the system.  For system size $N_0=4\times 4$
with $U=12t$, the contribution of the first basis state is about $80\%$. As
the system size reaching $N_0=10\times 10$ with $U=4t$, the contribution of
the first basis state is more than $90\%$.  As a consequence, for a same
accuracy requirement, the number of necessary basis states decreases with
system size.  The convergence rate is fast for the beginning several tens of
dominant basis states.  These basis states contribute more than $95\%$ to the
ground state for moderate correlation. Then contribution from each of the
following basis states drops rapidly. However, the major computation cost for
Fig.  \ref{fig1} is to search the remaining basis states.  In practical
calculations, one usually needs several tens of basis states for a reasonable
accuracy. 

We perform backward optimization for some cases to improve the accuracy.
There is very limited improvement from the backward optimization if the MP
method is converged. The improvement is usually less than $0.5\%$. If backward
optimization is performed before the convergence of MP process, the
improvement can be more than $1.5\%$ for some cases.

Our calculation is performed on single PC (AMD Opteron(tm) Processor 248).
Parallel implementation is easy for the current method. Since the major
numeric cost is the computation of the matrix elements of the Hamiltonian
during the search of the basis states, parallel implementation can be simply
realized by requiring each node handling some matrix elements of the
Hamiltonian.

There are many possible ways to improve the current method. For example, it is
worth to explore other type of basis states. In the present form, our method
is an extension to the mean field HF approximation. Mathematically, the
redundancy of the over-complete basis states is crucial for the convergence
speed of the MP algorithm~\cite{6}. By increase the redundancy of the
over-complete basis states, i.e., enlarging the searching space, it is possible
to speed up the convergence of MP method for searching the basis states. On
the other hand, for particles moving in 3D space, storage of a single particle
state needs sizable memory, one may choose the basis states for
single particles as product of one dimensional wave functions. In principle,
this method is able to compute the exited states. With some modifications, it
may be feasible to calculate the low-lying exited states.

In summary, the current method is stable and free from the sign problem. It
can apply to any system that can apply Hartree-Fock algorithm, and can be
regarded as an extension to the Hartree-Fock algorithm. Several tens of
determinant states are usually enough for meaningful result. This method may
offer an alternative to explore quantum effects of Many body systems.

This work is supported by the National Science Foundation of China (Grant No.
10375042). We acknowledge beneficial discussions with Prof. W.  Wang.

\end{document}